\documentclass[12pt]{article}
\usepackage{amsmath}
\usepackage{amssymb}
\usepackage{latexsym}
\usepackage{amsfonts,epsfig, mathrsfs}

\newcommand{\beq}{\begin{eqnarray}}
\newcommand{\eeq}{\end{eqnarray}}
\newcommand{\ber}{\begin{eqnarray}}
\newcommand{\eer}{\end{eqnarray}}

\newcommand{\C}{\mathbb{C}}

\def\bid{\hbox{1\hspace{-0.04in}I}} 

\usepackage[top=2.75cm, bottom=2.75cm, left=2.75cm, right=2.75cm]{geometry}

\begin{document}
\setcounter{page}{0}
\thispagestyle{empty}

\begin{flushright} \small
Imperial-TP-2009-CH-03
 \\UUITP-04/09\\
\end{flushright}
\smallskip
\begin{center} \LARGE
{\bf   Geometry of the
 $N=2$ supersymmetric sigma model with Euclidean worldsheet}
 \\[12mm] \normalsize
{\bf C.M.~Hull$^{b}$, U.~Lindstr\"om$^{a}$, L.~Melo dos
Santos$^{a,b}$, R. von Unge$^{c}$,
and M.~Zabzine$^{a}$} \\[8mm]
 {\small\it
$^a$Department of Physics and Astronomy\\
Uppsala University\\ Box 516, SE-75120 Uppsala, Sweden\\
 ~\\
$^b$The Blackett Laboratory\\ Imperial College\\
Prince Consort Road, London SW7 2AZ, U.K.\\
~\\
$^c$Institute for Theoretical Physics\\ Masaryk University\\
61137 Brno, Czech Republic
\\~\\
}
\end{center}
\vspace{8mm} \centerline{\bfseries Abstract} \bigskip

\noindent We investigate the target space geometry of supersymmetric  sigma models  in two dimensions with Euclidean signature, and the  conditions for  $N=2$
 supersymmetry.
For a real action,
  the geometry for the $N=2$ model is not the generalized K\"ahler geometry that arises for   Lorentzian signature, but is
   an interesting modification of this  which is not a complex geometry.

\eject
\section{Introduction}§ 

In this paper we discuss  $N=2$ supersymmetric sigma  models in 2 dimensions with
 Euclidean signature.
  One such model arises when the usual Lorentzian signature
  $N=1$ model is Wick-rotated and then required to have additional non-manifest supersymmetries.
  In this case, the Wess-Zumino term is imaginary and the action complex.
   This model was studied in connection with topological theories in \cite{HLMUZ2}.
   Below we  briefly discuss the target space geometry in this case. The R-symmetry group is
   $SO(2)\times SO(1,1)$  \cite{HLMUZ, HLMUZ2}  allowing an A-twist in which the $SO(2)$ factor is twisted with the  2d rotation group $SO(2)$ but not a B-twist.
   In  \cite{HLMUZ2} we considered the complexification of this model with R-symmetry
   $SO(2,\C)\times SO(2,\C)$ allowing both an A-twist and a B-twist with the complexified Lorentz group, which is also
   $SO(2,\C)$.

The main result of the paper concerns the  Euclidean model
with real action and real WZ term.
 The analysis closely follows that of GHR, (Gates, Hull and Ro\v cek)   \cite{GHS} in the Lorentzian
case, i.e., we make an ansatz for the extra supersymmetries and find the constraints on the target space
 geometry that follow from closure of the algebra and invariance of the action.  We find a curious generalization of
    complex geometry,
    which has a complex tensor $J$ that satisfies $J^2=-1$ and has vanishing Nijenhuis tensor.
    By complex tensor, we mean that it has components in a real coordinate system that are complex, whereas for a
    complex structure, the components would be real.
    We briefly discuss the underlying geometry.

We give the $N=2$ superspace formulation for
   the case in which the supersymmetry algebra closes off-shell.  In this case, the target space geometry has a metric of indefinite signature and
  two Yano f-structures \cite{Yano}.


\section{Sigma models}

The two-dimensional nonlinear sigma model has the   action
\begin{equation}
S=-\frac{1}{4}\int\limits_{\Sigma}d^2\sigma\,  \sqrt {h} \,
[h^{\mu\nu}\partial_\mu\phi^i\partial_\nu\phi^jg_{ij}(\phi)+\epsilon^{\mu\nu}\partial_\mu\phi^i\partial_\nu\phi^jB_{ij}(\phi)]~,
\label{lorentz bosonic action}
\end{equation}
for maps $\{\phi\}$ from a two dimensional manifold $\Sigma$ to a $d$-dimensional target space $M$:
\beq
\phi :~\Sigma \rightarrow M~.
\eeq
specified locally by functions $\phi^i (\sigma)$ giving the dependence of the real coordinates $\phi^i$ of $M$ on the real coordinates $\sigma ^\mu$ of $\Sigma$.
 The target manifold $M$ has a metric $g$ and 2-form potential $B$, while $\Sigma$ has a  metric $h_{\mu \nu}$ with $h=\vert \det (h_{\mu \nu})\vert$.
 The potential $B$ need only be locally defined, but there is a globally-defined closed 3-form field strength
 $H$ such that locally $H=dB$.
 The equations of motion depend on $B$ only through the
    3-form field strength $H$ and so are   well-defined.

  In  the usual case, the metric $h_{\mu \nu}$ has Lorentzian
 signature and $g_{ij}(\phi)$ and $B_{ij}(\phi)$ have real components.
The Euclidean version of this used in the path integral (given by a
Wick rotation in the case in which $h_{\mu \nu}$  is flat) is
 \begin{equation}
S=-\frac{1}{4}\int\limits_{\Sigma}d^2\sigma\,  \sqrt {h} \,
[h^{\mu\nu}\partial_\mu\phi^i\partial_\nu\phi^jg_{ij}(\phi)+i
\epsilon^{\mu\nu}\partial_\mu\phi^i\partial_\nu\phi^jB_{ij}(\phi)]~,
\label{Wick bosonic action}
\end{equation}
with $h_{\mu \nu}$ a Euclidean signature metric.
Note that the term involving $B$ is now pure imaginary, so that the action is complex.
For both the Lorentzian and Wick-rotated case, the quantum theory is well-defined if
$H$ is a globally-defined 3-form that represents an integral cohomology class, $H\in
H^3(\mathbb{Z})$. Geometrically this means that
there is a gerbe with curvature $H$ and connection $B_\alpha$ in each coordinate patch ${\cal O}_\alpha$. For the path integral, if
$H_2(M)$  is non-trivial, it is not sufficient
to specify $H$ , and a choice of $B$ must be made. Then the term containing the $B$-field
\beq\label{WZ}
e^{2\pi i\int\phi^*(B)}
\eeq
 defines the holonomy of a gerbe over the embedding of the world sheet. For
further details on gerbes and gerbe holonomy see \cite{hitchingerbe}, \cite{Hull:2008vw}.

For Euclidean signature one can also consider the real action
   (\ref{lorentz bosonic action}) with $h_{\mu \nu}$ a Euclidean signature metric. For the action to be well-defined,
   $B$ should be a globally-defined 2-form.
   However, the field equations are well-defined provided only that $H$ is a well-defined 3-form, so that a classical theory exists for any closed 3-form $H$.

 This paper will investigate the $N=2$ supersymmetrisations of both the real action
    (\ref{lorentz bosonic action}) and the complex action
    (\ref{Wick bosonic action}) for Euclidean $h_{\mu \nu}$. The motivation for this comes from our investigation of topological twistings  \cite{HLMUZ2}, where both cases played a role.

  An  $N=1$ supersymmetric version of these sigma model are obtained by promoting the $\phi$'s to
     $N=1$  superfields $\Phi(\sigma,\theta)$
  depending on  fermionic coordinates $\theta^\pm$, where $\theta^+$ has positive chirality and $\theta^-$ has negative chirality. In Lorentzian signature, $\theta^\pm$  are independent real Majorana-Weyl spinors, while in  Euclidean signature they are complex conjugate Weyl spinors, $(\theta^+)^*=\theta^-$.
  The corresponding supercovariant spinor derivatives  are $D_\pm$;
  see  \cite{HLMUZ, HLMUZ2}   for further discussion of our conventions.

  For both the Lorentzian sigma model with action  (\ref{lorentz bosonic action})  and Euclidean
  sigma model with complex action (\ref{Wick bosonic action}), the supersymmetric
  action is (taking $h$ to be flat)
\beq
S=-\frac{1}{4}\int d^2\sigma d^2\theta \, (
D_+\Phi^iE_{ij}(\Phi)D_-\Phi^j)~,
\label{oneone action}
\eeq
where
\beq\label{eij}
E_{ij}=g_{ij}+B_{ij}~,
\eeq
By contrast, for the Euclidean
  sigma model with real action (\ref{lorentz bosonic action}), the $N=1$ supersymmetric version is again given
  by (\ref{oneone action}) but now
    \beq\label{eijcom}
E_{ij}=g_{ij}+iB_{ij}~,
\eeq
is complex.

For special target space geometries, these $N=1$ sigma models can have extra supersymmetries. For example, the Lorentzian sigma model will have $N=2$ supersymmetry provided the target space has
the bihermitean geometry of GHR \cite{GHS}, which has recently been given a new formulation in terms of Generalized K\"ahler geometry \cite{gualtieriPhD},
\cite{Lindstrom:2005zr}.
Here
we will examine the geometries needed for the real and Wick-rotated $N=1$ Euclidean sigma models to have $N=2$ supersymmetry.

\section{Geometry of the classical models }
\label{class1}

In this section we   briefly review the geometric structure of the target spaces for the Lorentzian
 and Wick-rotated models. We then present our main results that concern
 the geometry for the    Euclidean model with real action.

\subsection{The Lorentzian   $N=2$ model}
\label{Lorentzian}

We start with the Lorentzian signature $N=1$ supersymmetric action
(\ref{oneone action}) with (\ref{eij}) and follow the analysis of \cite{GHS}.
The general ansatz for an extra right and left supersymmetry is
\begin{equation}
\delta_\epsilon \Phi^i= i{J_+}^i_{\phantom{i}j}(\epsilon_-D_+\Phi^j) +i{J_-}^i_{\phantom{i}j}(\epsilon_+D_-\Phi^j)~,
\label{second transformation}
\end{equation}
where $\epsilon_\pm$ are independent real supersymmetry transformation parameters and $J_\pm$ are some mixed real  tensors on $M$.
Closure of the supersymmetry algebra and invariance of the action then impose conditions on $J_\pm$. Closure  requires that
$J_\pm$ are complex structures,
\beq
J_\pm^2=-1~, ~~{\cal N}(J_\pm)=0~,
\label{cst}
\eeq
where ${\cal N}(J)$ denotes the Nijenhuis tensor.
Invariance of the action requires that
they are also covariantly constant with respect to connections with torsion,
\beq
\nabla^\pm J_\pm=0~,
\label{int}
\eeq
and the metric $g$ is hermitean with respect to both
\beq
J_\pm^tgJ_\pm=g~.
\label{presg}
\eeq
The connections with torsion are constructed from the Levi-Civita connection $\Gamma$ and the 3-form   $H=dB$:
\beq
\Gamma^\pm=\Gamma \pm \frac 1 2 g^{-1}H~.
\label{torsion}
\eeq
Then $M$ has a GHR bihermitian geometry   \cite{GHS}.

\subsection{The   Wick rotated $N=2$ model}
\label{Wicked}

Consider next the \lq Wick-rotated' model
given by $N=1$ supersymmetric action
(\ref{oneone action}) with (\ref{eij}) and Euclidean world-sheet metric, so that the component expansion
has   bosonic part
    (\ref{Wick bosonic action}) with imaginary WZ term.
The anasatz for the extra supersymmetry is again
(\ref{second transformation}) but now all spinors are complex, with
\beq
(\epsilon_\pm)^*=\epsilon_\mp~,~~(D_\pm)^*=D_\mp~.
\label{conjugat22}
\eeq
The algebra of the supercovariant derivatives is
\begin{eqnarray}
\{D_+,D_+\}&=&\partial~,\nonumber\\
\{D_-,D_-\}&=&\bar{\partial}~,
\label{esusy}
\end{eqnarray}
where the partial derivatives on the right are derivatives with respect to  $z=\sigma ^1+i \sigma ^2$ and $\bar z=\sigma ^1-i \sigma ^2$ respectively.
Closure of the algebra and invariance of the action
give  the same set of equations  (\ref{cst})-(\ref{torsion}) as for the Lorentzian case.
However, the reality conditions on $\Phi$ and the transformations (\ref{second transformation}) give us the condition
\begin{equation}
J_+^*=J_-~.
\label{complex conjugation of J}
\end{equation}
The complex conjugate of (\ref{int})
now yields
\begin{equation}
\nabla^{\pm}J_{\mp}=0~,
\end{equation}
which together with (\ref{int}) implies the K\"ahler equation,
\begin{equation}
\nabla J_{\pm}=0~,
\end{equation}
and
\begin{equation}
H=0~.
\end{equation}
 Indeed this should not come as surprise. The Wick-rotated action is complex and
 so the real and imaginary parts must be separately invariant,
 so that the geometry must be K\"ahler and the WZ term trivial.
 Then $B$ is a connection on a flat gerbe and the expression  $e^{2\pi i\int\phi^*(B)}$ is well-defined
and gives us the holonomy of the flat gerbe.

\subsection{The Euclidean $N=2$ model  with real action }
\label{oneone}

We now consider the action (\ref{oneone action})
with Euclidean world-sheet and
\beq
E_{ij}=g_{ij} + iB_{ij}~,
\eeq
so that the component expansion has the bosonic term   (\ref{lorentz bosonic action}) with real
  $B$-field term.
 The anasatz for the extra supersymmetry is again
(\ref{second transformation}) but now all spinors are complex, with
 the reality conditions on $\Phi$ and the transformations (\ref{second transformation})  again giving   $
J_+^*=J_-$.
Then  the second supersymmetry variation (\ref{second transformation})
    becomes
\begin{equation}
\delta_\epsilon \Phi^i= i{J}^i_{\phantom{i}j}(\epsilon_-D_+\Phi^j) +i{J^*}^i_{\phantom{i}j}(\epsilon_+D_-\Phi^j)~,
\label{second transformationreal}
\end{equation}
 where $J^*$ is the complex conjugate to $J=J_+$.
Alternatively we can split $J$ into real and imaginary parts
\beq
J=f + i\tilde f ~,
 \label{J expanded}
\eeq
where $f$ and $\tilde f$ are real tensors, so that the transformation
(\ref{second transformationreal}) becomes
\begin{equation}
\delta_\epsilon \Phi^i=if^i_{~j}(\epsilon_-D_++\epsilon_+D_-)\Phi^j+\tilde f^i_{\phantom{i}j}(\epsilon_+D_--\epsilon_-D_+)\Phi^j~.
\label{ftransf}
\end{equation}

The conditions for supersymmetry following \cite{GHS} are similar to before, but with extra factors of $i$.
 The on-shell closure of the  supersymmetry algebra  implies that
 $$J^2=-1$$ and its Nijenhuis tensor vanishes, ${\cal N}(J)=0$.
 Invariance of the action under the second supersymmetry (\ref{second transformation}) requires
the metric $g$ must satisfy
\beq
  J^t g J = g
 \label{defindkslslw}
 \eeq
together with
\begin{equation}
\nabla_H J =0 ~,
\label{covariance eq}
\end{equation}
 where $\nabla_H$ has connection
 $$\Gamma _H=\Gamma + \frac i 2 g^{-1}H~.$$
   Thus formally our new conditions are similar to the generalized K\"ahler geometry, but
  now $J_+ = J$, $J_- = J^*$ are complex tensors and
the torsion term in the connection now has a    factor of $i$.
Thus the conditions are formally similar to those for generalized K\"ahler geometry,  but the different reality properties and extra factors of $i$
means that  the implications of these conditions will be quite different.

The target manifold $M$ is not a complex manifold in the standard sense. We still can define the projectors
\begin{equation}
p_\pm=\frac{1}{2}(1\pm iJ)~,~~~~~~~~p^*_\pm=\frac{1}{2}(1\mp iJ^*)~,
\end{equation}
  which would give us four integrable complex  distributions on the complexified tangent bundle $TM_{\mathbb C}$.
   However we would not be able to define the decomposition of a vector into holomorphic and antiholomorphic parts.
    For example, the projector
   $p_-$ would define the "holomorphic" vectors $T^{(1,0)}M$, but unlike the complex manifold we have now
   \beq
    T^{(1,0)}M \cap T^{(0,1)}M \neq \emptyset~,
   \eeq
 where $T^{(0,1)}M$ is the subbundle complex conjugate to $T^{(1,0)}M$.

 Using  the real and imaginary parts of $J$   introduced in (\ref{J
expanded}) the condition  $J^2=-1$ becomes
\begin{eqnarray}
f^2-\tilde f^2&=&-1~,\nonumber\\
\{f,\tilde f\}&=&0~.
\end{eqnarray}
In terms of real tensors,  the condition (\ref{covariance eq})
can be written as two real equations,
\begin{eqnarray}
\nabla f&=&\frac 1 2 g^{-1}H\tilde{f}~,\nonumber\\
\nabla\tilde{f}&=&-\frac 1 2 g^{-1}Hf~.
\end{eqnarray}
 Furthermore, as in the generalized K\"ahler case \cite{Lyakovich}
  we can define two real Poisson structures
\ber
\label{poisson1}\pi_+&= &\frac{1}{2} (J +J^*)g^{-1}= f g^{-1}~,\\
\label{poisson2}\pi_-&=& \frac{1}{2i} (J -J^*)g^{-1}= \tilde f g^{-1}~,
\label{poisson structures}
\eer
 which   define   symplectic foliations.  Locally we can choose the coordinates adapted to these foliations
  and $f$, $\tilde{f}$   look relatively simple in those coordinates.

\subsection{Off-shell closure and f-structures}

The $N=2$ superalgebra will close off-shell only if
 $J$ and $J^*$  commute\footnote{If
  auxiliary fields are included,   the situation changes. See \cite{Lindstrom:2005zr}.}.
 In this case the condition  $[J,J^*]=0$ becomes
\begin{equation}
f\tilde f=0~.
\label{sector condition}
\end{equation}
Then at least one of the two structures $f, \tilde f$ must be degenerate.
Then
$f$ and $\tilde{f}$ satisfy
\begin{eqnarray}
f^3+f&=&0~,\label{f eq}\\
\tilde f^3-\tilde f&=&0~.
\label{f tilde eq}
\end{eqnarray}
Equation (\ref{f eq}) is the generalization of an almost
 complex structure condition ($f^2=-1$)  to  allow   the possibility of $f$ being a degenerate tensor.
A tensor $f$ of constant rank satisfying   (\ref{f eq}) is a
 {\em{ Yano f-structure }}
  \cite{Yano}.
Similarly, equation (\ref{f tilde eq}) is the generalization of an almost  product structure ($\tilde{f}^2=1$) condition with $\tilde{f}$ being possibly degenerate and gives a generalised f-structure.

Let
\begin{equation}
P= - f^2~.
\end{equation}
Then
\begin{equation}
 P^2 =P~,
\end{equation}
so that $P$ is a projector.
At a point, if the rank of $P$ is $r$, then
we can choose a basis in which $P$ has a block form
\begin{equation}
P=\left[\begin{array}{c|c} {\bf 0} & \\
  \hline
 &\bid   \\
 \end{array}\right]~,
\end{equation}
where $\bid $ is the $r\times r$ unit matrix and
 ${\bf 0}  $ is the $(D-r)\times (D-r)$ zero matrix.
Then
	\begin{equation}
f=\left[\begin{array}{c|c} {\bf 0} & \\
  \hline
 &j   \\
 \end{array}\right]~,
\end{equation}
 where
 $j$ is an $r\times r$ non -degenerate matrix satisfying
 $$j^2=-\bid~.
 $$
 This implies that $r$ is even, $r=2q$, and one can choose a basis so that
\begin{equation}
j=\left(\begin{array}{cc} {  0} &\bid \\
- \bid &0    \\
 \end{array}\right)~.
\end{equation}

Next, since $f\tilde f =0$,  $\tilde f $ has the block form
\begin{equation}
\tilde{f}=\left[\begin{array}{c|c}  \pi & \\
  \hline
 &{\bf 0}   \\
 \end{array}\right]~,
\end{equation}
where $\pi $ is a $(D-r)\times (D-r)$ matrix satisfying
$$\pi ^3 -\pi =0~.
$$
Then $\pi$ has eigenvalues $\pm 1,0$ and take the form
\begin{equation}
\left(
\begin{array}{ccc}
\bid  &  {\bf 0}  & {\bf 0}    \\
{\bf 0}   &- \bid  & {\bf 0}   \\
 {\bf 0}  & {\bf 0}   & {\bf 0}
\end{array}
\right)
\end{equation}
split into blocks of dimension $a,b,c$ with $a+b+c+r=D$.
If the number $c$
of zero eigenvalues is non-zero, then there will be a subspace on which the 2nd supersymmetry does not act. If it is non-degenerate, then
$$\pi^2=\bid
$$ and we will mostly be interested in this non-degenerate case.

Finally
	\begin{equation}
J=\left[\begin{array}{c|c} i \pi &0 \\
  \hline
 0&j   \\
 \end{array}\right]~.
 \label{Jsplit}
\end{equation}
The vanishing of the Nijenhuis tensor for $J$ implies that we can choose coordinates so that $J$ takes this form on a patch.

As in Lorentzian signature \cite{GHS}, we can (for non-degenerate $\pi$)
define a local \emph{product structure}  defined by a real tensor $\Pi$,
\begin{equation}
\Pi=JJ^*
\label{Jproduct}
\end{equation}
satisfying
$$\Pi ^2=\bid~,
$$
which takes the form
\begin{equation}
\label{Pisplit}
\Pi=\left[\begin{array}{c|c}\bid &0 \\
  \hline
 0&-\bid \\
 \end{array}\right]
\end{equation}
 and this local product structure $\Pi$ is integrable.

We will see in the next section that $N=2$ superspace naturally gives geometries with $a=b=p,c=0$
and which admit a local description in terms of single real function, very much in analogy with the K\"ahler and generalized K\"ahler cases.


\subsection{The Euclidean $N=2$ model  with real action in $N=2$ superspace}

The $N=1$-supersymmetric action (\ref{oneone action}) with an
extra supersymmetry (\ref{second transformation}) that closes off-shell\footnote{In  Lorentzian signature, a complete description covering all off-shell cases requires additional $N=(2,2)$ semi-chiral fields.} can be
reformulated in Euclidean $N=2$ superspace:
\begin{equation}
S=2\int d^2z\, d^2 \theta \, d^2 \bar \theta \, \,
   K(\phi, \bar{\phi}, \chi, \tilde{\chi})~.
\label{22 action}
\end{equation}
 We remind the reader that in Euclidean signature, the conjugation relations for $N=2$ spinor derivatives are 
 \beq
 (\mathbb{D}_\pm)^\dagger= \bar {\mathbb{D}}_\mp~.
 \label{ebat}
 \eeq
In (\ref{22 action})  $\phi^u\quad u=1,...,p$ are chiral superfields
($\bar{\mathbb{D}}_+\phi=\bar{\mathbb{D}}_-\phi=0$) and $\bar{\phi}^u$ their complex
conjugates ($\mathbb{D}_+\bar{\phi}=\mathbb{D}_-\bar{\phi}=0$).
The   fields $\chi^a$
are mixed chiral fields  satisfying
$$\bar{\mathbb{D}}_+\chi=\mathbb{D}_-\chi=0$$ where $ a=1,...,q$. In  Euclidean signature
its complex conjugated field $\bar{\chi}$ is still a mixed
chiral field, as
\begin{equation} \bar{\mathbb{D}}_+\bar{\chi}=\mathbb{D}_-\bar{\chi}=0
\end{equation}
so there is no loss of generality in taking the fields $\chi^a$ to be real.
The fields $\tilde \chi^a$ are
real mixed anti-chiral  fields   satisfying
\beq
\mathbb{D}_+\tilde{\chi}=\bar{\mathbb{D}}_-\tilde{\chi}=0~.
\eeq
We write the action in terms of
 $q$
real mixed chiral fields $\chi^a$ and an equal number\footnote{A different number of mixed chiral and mixed anti-chiral fields leads to degenerate models.} of real mixed anti-chiral  fields $\tilde \chi^a$.
 This structure of the potential $K$ was first introduced and discussed in
\cite{HLMUZ}.

We now relate this to the
action in Sec.\ref{oneone}. Note that the bosonic part of the action is    (\ref{lorentz bosonic action}),
 which can be written using
complex world-sheet coordinates $z,\bar z$ as
\begin{equation}
S=-\int\limits_\Sigma d^2z( g_{ij}\partial
\phi^i\bar{\partial}\phi^j-iB_{ij}\partial\phi^i\bar{\partial}\phi^j)~.
\label{complex action}
\end{equation}

We write the action (\ref{22 action}) as
\begin{equation}
S=\int
d^2z
(\mathbb{D}_-\bar{\mathbb{D}}_-\mathbb{D}_+\bar{\mathbb{D}}_++\bar{\mathbb{D}}_-\mathbb{D}_-\bar{\mathbb{D}}_+\mathbb{D}_+)K(\phi,\bar{\phi},\chi,\tilde{\chi})\Big\vert~,
\end{equation}
where $(..)\Big\vert$ denotes taking the $\theta =0$ part.
The bosonic part of the action then becomes
\begin{equation}
\label{expanded}
\begin{split}
S=\int
d^2z\,
 \Bigl(&
-K,_{\bar{u}v}\partial\phi^v\bar{\partial}\bar{\phi}^u-K,_{u\bar{v}}\partial\bar{\phi}^v\bar{\partial}\phi^u
\\
&+K,_{\tilde{a}b}\partial\chi^b\bar{\partial}\tilde{\chi}^a+K,_{a\tilde{b}}\partial\tilde{\chi}^b\bar{\partial}\chi^a
\\
&+K,_{u\bar{v}}\partial\bar{\phi}^v\bar{\partial}\phi^u-K,_{\bar{u}v}\partial\phi^v\bar{\partial}\bar{\phi}^u\\
&+K,_{ua}\partial\chi^a\bar{\partial}\phi^u-K,_{au}\partial\phi^u\bar{\partial}\chi^a
\\
&+K,_{a\bar{u}}\partial\bar{\phi}^u\bar{\partial}\chi^a-K,_{\bar{u}a}\partial\chi^a\bar{\partial}\bar{\phi}^u \Bigr)~.
\end{split}
\end{equation}

Comparing the  actions
(\ref{complex action}) and
(\ref{expanded}), we learn about
the geometry of the target space
manifold.
The metric $g$   has a block diagonal structure,
\begin{equation}
\mathbf{g}=\left[\begin{array}{cc|cc} 0&K,_{u\bar{v}}&&\\
K,_{\bar{u}v}&0&&\\
\hline
&&0&-K,_{a\tilde{b}}\\
&&-K,_{\tilde{a}b}&0\end{array}\right]~,
\label{sdfdsf}
\end{equation}
where we have a block with $2p\times 2p$ entries for the chiral
sector and a block of $2q\times2 q$ for the mixed chiral
sector.
The chiral sector block of dimension $2p$ has
  Euclidean signature while the  mixed chiral sector block of dimension $2q$ has a metric of split signature
 $(q,q)$  with $q$ positive
eigenvalues and $q$ negative ones.

The 2-form $B$ has off-diagonal blocks mixing
chiral with mixed chiral derivatives plus an extra bloc
for the chiral sector,
\begin{equation}
\mathbf{B}=\left[\begin{array}{cc|cc}
0&-iK,_{u\bar{v}}&-iK,_{ua}&0\\
iK,_{\bar{u}v}&0&iK,_{\bar{u}a}&0\\
\hline
iK,_{au}&-iK,_{a\bar{u}}&&\\
0&0&&\end{array}\right]~.
\end{equation}
It has a different form from that in the standard GHR-gauge \cite{GHS}, which gives a $B$-field that
  is complex in Euclidean signature. Here we use an alternative
gauge in which $B$ is real when written in real coordinates.

We now turn to   the   structures
$J$ and $J^*$ that appear in the   supersymmetry transformations
(\ref{second transformationreal}), following  \cite{GHS}.
The
 $N=2$ superspace formulation makes the extra supersymmetry manifest.
 Expanding into $N=1$ superfields gives transformations of the form (\ref{second transformationreal})
 and from these one can read off the  structures
$J$ and $J^*$, which are constant in this coordinate system.
We define the Weyl $N=1$ spinor derivative $D_\pm$, and the generator of the non-manifest supersymmetry $Q_\pm$,
\begin{eqnarray}
D_\pm&=&\frac{1}{\sqrt{2}}(\mathbb{D}_\pm+\bar{\mathbb{D}}_\pm)~,\\
Q_\pm&=& \frac{i}{\sqrt{2}} (\mathbb{D}_\pm-\bar{\mathbb{D}}_\pm)~.\nonumber
\end{eqnarray}
 The $N=1$ algebra with the property (\ref{conjugat22}) and $\{D_\pm , Q_\pm \}=0$
    follow from the $N=2$ algebra  and the  property (\ref{ebat}). 
The $Q$-transformations of the $N=1$  fields ($\phi$, $\bar{\phi}$, $\chi$, $\tilde{\chi}$)  are
\begin{eqnarray}
\delta_\epsilon \phi^u&=&i\epsilon_-Q_+\phi^u+i\epsilon_+Q_-\phi^u\nonumber\\
&=&-\epsilon_-D_+\phi^u-\epsilon_+D_-\phi^u~,\nonumber\\
\delta_\epsilon \bar{\phi}^u&=&\epsilon_-D_+\bar{\phi}^u+\epsilon_+D_-\bar{\phi}^u~,\\
\delta_\epsilon \chi^a&=&-\epsilon_-D_+\chi^a+\epsilon_+D_-\chi^a~,\nonumber\\
\delta_\epsilon \tilde{\chi}^a&=&\epsilon_-D_+\tilde{\chi}^a-\epsilon_+D_-\tilde{\chi}^a~.\nonumber
\label{N1fexpansion}
\end{eqnarray}
To relate this to  the structure that we found for $J$ in the previous section we need to expand the real $N=1$ superfield $\Phi$ in real components, so we need to split the $N=2$ chiral superfield $\phi$ and its antichiral partner $\bar{\phi}$ into their real components,
$$\phi =\phi_1+i\phi_2~,$$
  $$\bar{\phi}=\phi_1-i\phi_2~.$$
 Writing the $N=1$ superfields $\Phi$ in terms of the  real $N=1$ superfields ($\chi$, $\tilde{\chi}$, $\phi_1$, $\phi_2$), 
   we can read off   the $J$ in transformation (\ref{second transformationreal}) to be 
\begin{equation}\label{Jkkkkk}
J=\left[\begin{array}{cc|cc} i&0&&\\
0& -i&&\\
\hline
&&0& -1\\
&&1&0 \end{array}\right]~.
\end{equation}
 We thus recover the structures discussed in the previous subsection, cf. (\ref{Jsplit})-(\ref{Pisplit}).

\section{Conclusions}

As discussed, e.g., in \cite{HLMUZ},\cite{Buscher:1987uw},\cite{AbouZeid:1999em}, Euclidean supersymmetry differs in many ways from the usual Lorentzian one. In this article this is again illustrated by considering the target space geometry of a ``natural'' sigma model
in Euclidean signature.  We encountered the modification of the complex geometry defined by the complex tensor $J$ which
 formally satisfies the usual definitions of complex structure, but is  now a complex tensor.  We considered in detail
  the geometry that emerges from off-shell supersymmetry which differers from the usual case
 both in the signature of the metric and in the additional structure it carries.
The geometry is described by $(M, g, B, f, \tilde f)$ where $f$ and $\tilde f$ are Yano f-structures when the superymmetry algebra closes off-shell.
This structure is derived from a potential  as in the Lorentzian case.
The field equations are well-defined provided only that $H$ is globally defined, but the quantum theory requires further that $H$ represent a trivial cohomology class.
It is only the special case of a K\"ahler manifold as target space that can be described by $N=2$-supersymmetric models of all three kinds discussed here.
\vspace{2cm}

\noindent
{\large \bf Acknowledgement:}\\

The research of UL was supported by EU grant (Superstring theory) MRTN-2004-512194 and VR grant 621-
2006-3365.
The research of R.v.U. was supported by the Simons Center for Geometry and Physics as
well as the Czech ministry of education under contract No. MSM0021622409. The research
of M.Z. was supported by VR-grant  621-2008-4273.
All the authors are  happy to thank the program ``Geometrical Aspects of String Theory'' at Nordita,
where part of this work was carried out.

\end{document}